\documentclass[prb,%
 reprint,
 amsmath,amssymb,
 aps,
]{revtex4-2}

\usepackage{physics}
\usepackage{amsmath,amssymb,amsfonts}
\usepackage{comment}
\usepackage{subcaption}
\usepackage{xcolor}
\usepackage{graphicx}
\usepackage{epstopdf} 
\usepackage{dcolumn}
\usepackage{bm}

\usepackage{hyperref}
\hypersetup{colorlinks = true, 
	    linkcolor = blue, 
	    urlcolor = blue,
            citecolor = blue}
\newcommand*{\dif}{\mathop{}\!\mathrm{d}}
\newcommand*{\pd}{\mathop{}\!\partial}
\newcommand*{\D}{\mathop{}\!\mathcal{D}}

\newcommand*{\ovl}{\mathop{}\!\overline}

\def\*#1{\mathbf{#1}} 
\def\+#1{#1^{\dagger}}
\def\$#1{\mathcal{#1}}
\def\-#1{\bar{#1}}
\def\^#1{\hat{#1}}
\begin{document}

\preprint{APS/123-QED}

\title{A Replica Stoner Theory for Dirty Ferromagnets}

\author{Wenzhe Deng}
\author{Tai Kai Ng}%
 \email{phtai@ust.hk}
\affiliation{%
 Department of Physics, Hong Kong University of Science and Technology, Clear Water Bay, Kowloon, Hong Kong
}%

\date{\today}

\begin{abstract}
This paper investigates the effect of disorder on a ferromagnetic metal with repulsive interactions. We assume that, in the clean limit, the ferromagnetic state can be described by Stoner mean-field theory and study how disorder affects the the system by using a combined replica + Stoner mean-field approach. At zero temperature, we find that a replica-symmetric ferromagnetic mean-field solution exists in the presence of disorder with a modified Stoner criteria where the ferromagnetism is enhanced by disorder. At finite temperature, a Landau theory is employed to construct the phase diagram, revealing that beyond a critical disorder strength, a spin-glass phase may exist between the high-temperature paramagnetic phase and the low-temperature ferromagnetic phase. For weak (repulsive) interaction where the system is non-ferromagnetic in the clean limit, the possibility of a disordered-induced ferromagnetic ground state is observed both at zero temperature and finite temperature. The potential applicability of this framework to realistic materials is briefly discussed.

\end{abstract}

\maketitle


\section{\label{Sec:intro}Introduction}
The influence of disorder on magnetic systems has garnered significant interest within both experimental and theoretical physics communities \cite{General_EXP1,General_EXP2,General_EXP3,General_THE1,General_THE2,General_THE3,General_THE4}. In particular, the study of itinerant electron magnetism in the presence of disorder has been extensively explored through various approaches, including diagrammatic perturbation techniques \cite{Itin_THE1(perturbation),Itin_THE5(Diagram)}, mean-field theories \cite{Itin_THE2(MF)}, Monte-Carlo simulations \cite{Itin_THE2(MC)}, and self-consistent first-principle numerical calculations \cite{Itin_THE3(FP),Itin_THE4(FP),Itin_THE6(DMFT)}. These studies have provided us understandings on the stability, phase transitions, and transport phenomena in disordered ferromagnets. Experimental investigations have also revealed intriguing properties of ferromagnetic materials subjected to different types of disorder \cite{Itin_EXP1,Itin_EXP2,Itin_EXP3,Itin_EXP4}. Notably, some experiments suggest that disorder can enhance ferromagnetism across a range of temperatures, while others indicate that strong disorder suppresses magnetic order \cite{Itin_EXP1}. 

In this work, we propose a semi-quantitative theory of ferromagnetism in a dirty metal. Specifically, we focus on systems  where ferromagnetism can be described by Stoner mean-field theory \cite{Stoner_ori} in the clean limit. For dirty metals, we combine the Stoner mean-field theory and replica-trick to study the effect of disorder on Stoner ferromagnetism. Our microscopic model considers electrons on a lattice with a random on-site potential and on-site repulsive interactions. Disorder effects are encoded in two order parameters within the mean-field equations derived from the replica-Stoner approach. We shall first consider the zero-temperature mean-field solutions in our theory and then construct the corresponding finite temperature Landau free energy to study the finite temperature magnetic phase transitions. Our results indicate that disorder can enhance ferromagnetism in ferromagnetic materials and even induce magnetism in systems which are non-magnetic in the clean limit. We also find that beyond a critical disorder strength, a spin-glass phase may exist between the high-temperature paramagnetic phase and the low-temperature ferromagnetic phase. These findings qualitatively align with recent experimental studies \cite{Itin_EXP2,Itin_EXP3,Itin_EXP4} and theoretical studies \cite{Itin_THE1(perturbation),Itin_THE2(MF),Itin_THE6(DMFT)}. The framework presented here, combining mean-field theory and replica-trick, can be extended to other systems such as dirty superconductors \cite{Replica_BCS} and spin-liquid states.

\section{\label{the_model}Formulation}
We begin with a Hamiltonian $\$H = \$H_0 + \$H_{\rm{int}}$ describing interacting electrons on a lattice with random lattice potential and a Hubbard-type interaction:
\begin{gather}
    H_0 = \sum_{ij,\sigma} t_{ij}\+c_{i\sigma}c_{j\sigma} + \sum_{i,\sigma}W_in_{i\sigma}, \notag \\
    H_{\mathrm{int}} = \sum_{i} U \+c_{i\uparrow}c_{i\uparrow}\+c_{i\downarrow}c_{i\downarrow} \label{original_model}
\end{gather}
where $\+c_{i\sigma}(c_{i\sigma})$ is the electron creation (annihilation) operator at site $i$. $W_i$ is a random on-site potential and $n_{i\sigma} = \+c_{i\sigma}c_{i\sigma}$. We focus on the case that the on-site interaction is repulsive ($U>0$) and the ferromagnetism is described by a Hartree mean-field theory which is essentially Stoner theory.

The mean-field theory is most conveniently performed by writing the interaction parts of the Hamiltonian $\$H_{\rm{int}}$ as the sum of spin-spin interactions and the density-density interactions:
\begin{equation}
    \sum_i Un_{i\uparrow}n_{i\downarrow} = \sum_i -\frac{1}{3}U\*S_i\cdot \*S_i + \sum_i \frac{U}{4}n_in_i,
\end{equation}
in which $n_i = n_{i\uparrow} + n_{i\downarrow}$ and $\*S_i = \frac{1}{2}\+c_{i\alpha}\bm\sigma_{\alpha\beta}c_{i\beta}$ are the density and spin operator on site $i$, respectively, with $\bm\sigma_{\alpha\beta}=(\sigma^x_{\alpha\beta}, \sigma^y_{\alpha\beta}, \sigma^z_{\alpha\beta})$ representing the Pauli matrices. As we are interested only in magnetism, we assume that the density-density interaction gives rise only to a modified random potential $W_i\rightarrow \overline{W}_i=W_i+U\expval{n_i}/4$ which will be neglected in the following, i.e., we replace
\begin{equation}
    \sum_i Un_{i\uparrow}n_{i\downarrow} \rightarrow \sum_i -\frac{1}{3}U\*S_i\cdot \*S_i.
\end{equation}
We shall absorb the factor $\frac{1}{3}$ into definition of $U$ in the following.

In the eigenstate basis given by $[c_{k\sigma},H_0-\mu N]=\xi_k c_{k\sigma}$
where $N$ is the total particle number operator and $c_{k\sigma}=\sum_i\phi_{k}(\*r_i)c_{i\sigma}$ where $\phi_{ki}$ are the eigenstates of $H_0$, (we set $\hbar=1$), the Hamiltonian can be written as (see also ref.~\cite{Disorder_SC})
\begin{equation}
H=\sum_{k \sigma} \xi_k c_{k \sigma}^{\dagger} c_{k \sigma}-\sum_{k q, l p} U_{k q, l p} \*S_{kq} \cdot \*S_{lp},
\end{equation}
where $\xi_k$ is the single-particle energy of state $k$ measured from the chemical potential $\mu$, $\*S_{kq}=\frac{1}{2}\+c_{k\alpha}\bm\sigma_{\alpha\beta}c_{q\beta}$ and
\begin{equation}
U_{k q, l p}=\sum_i U \phi^*_{k}(\*r_i) \phi_{q}(\*r_i)\phi^*_{l}(\*r_i) \phi_{p}(\*r_i). 
\end{equation}
We note that $k$ represents eigenstates of $H_0$ which are not plane-wave states. We shall assume that $\xi_k$ forms a continuous spectrum across the Fermi surface.

\subsection{Mean-Field Theory}
 We introduce the mean-field decoupling
\begin{align}
&\*S_{kq} \cdot \*S_{lp}  \rightarrow  \*S_{kq} \cdot \langle\*S_{lp}\rangle + \langle\*S_{kq}\rangle \cdot \*S_{lp}-\langle\*S_{kq}\rangle \cdot \langle\*S_{lp}\rangle \notag \\ 
 & \sim  \delta_{lp}\delta_{kq} \Big( \*S_{k} \cdot \expval{\*S_{p}} +\expval{\*S_{k}} \cdot \*S_{p}-\expval{\*S_{k}} \cdot \expval{\*S_{p}} \Big) \label{mf1}
\end{align}
where $\*S_{p}=\*S_{pp}$ and $\langle...\rangle$ denotes thermal average. Denoting $U_{kk, pp}\equiv U_{kp}$, the corresponding mean-field Hamiltonian is
\begin{equation}
\label{mf2}
    H_{MF}=\sum_{k \sigma} \xi_k c_{k \sigma}^{\dagger} c_{k \sigma}-\sum_{k} \left(2\*h_{k} \cdot \*S_{k}-\*h_{k} \cdot \langle \*S_{k}\rangle\right), 
 \end{equation}
where $\*h_{k}=\sum_{p(\neq k)}U_{kp}\langle\*S_{p}\rangle$. The corresponding eigen-energies are $\xi_{k\sigma} = \epsilon_k -\sigma|{\*h_k}|$, where $\sigma=\pm1$, denoting eigenstates $\phi_{k\sigma}(\*r_i)$ with spin parallel and anti-parallel to effective magnetic fields $\*h_k$.

To understand our mean-field theory, we compare it with the usual Hartree theory based on the mean-field decoupling
\begin{equation}
\*S_i \cdot \*S_i \sim 2\langle\*S_i\rangle \cdot \*S_i - \langle\*S_i\rangle \cdot \langle\*S_i\rangle.
\end{equation}
The corresponding eigenstates $\phi^{\mathrm{HF}}_{a}(\*r_i)$ with index $a=(k,\alpha=\pm1)$ satisfies the mean-field equation
\begin{widetext}
\begin{equation}
    \xi_{k\alpha}^{\mathrm{HF}}\phi^{\mathrm{HF}}_{k\alpha}(\*r_i)= \sum_{\beta,j}\big[t_{ij}\delta_{\alpha\beta} +\delta_{ij}\big(\overline{W}_i\delta_{\alpha\beta} - U\expval{\*S_i}\cdot\bm\sigma_{\alpha\beta}\big) \big]\phi^{\mathrm{HF}}_{k\beta}(\*r_j).
\end{equation}
\end{widetext}
To arrive at our mean-field theory we assume that disorder is strong enough so that all states $\phi^{\mathrm{HF}}_a(\*r_i)$ are localized. We also assume that $U\expval{\*S_i}$ remains more-or-less constant within the region where the state $\phi^{\mathrm{HF}}_a(\*r_i)$ exists, and we approximate: $U\expval{\*S_i}\sim {\*h_a}$ where $\*h_a$ represents mean value of $U\expval{\*S_i}$ in this region. Notice we do not assume that ${\*h_a}$ is the same for all states $\phi^{\mathrm{HF}}_a(\*r_i)$. Correspondingly
\begin{equation}
    \sum_i\phi_{k\alpha}^*(\*r_i)U\expval{\*S_i}\cdot\bm\sigma_{\alpha\beta}\phi_{k\beta}(\*r_i) \sim \bm h_a\cdot\langle\bm\sigma_{\alpha\beta}\rangle_a\label{HF_basic_approx}
\end{equation}
where $\langle\bm\sigma_{\alpha\beta}\rangle_a=\sum_i\phi^*_{k\alpha}(\*r_i)\bm\sigma_{\alpha\beta}\phi_{k\beta}(\*r_i)$. With this approximation, we recover our mean-field theory given in Eq.~\eqref{mf1} and Eq.~\eqref{mf2} after some simple algebra.

The above approximation yields a clean ferromagnet when ${\*h_k}\sim {\*h}$ is constant as in disorder-free systems. For localized electronic states, we expect that fluctuation in ${\*h_k}$ exists because of randomized coupling between localized states that are far apart. For states $k$ close to the Fermi surface with $|{\*h_k}|>\xi_k$, the state $(k{\downarrow})$ becomes unoccupied resulting in formation of a local magnetic momentum polarized along the direction of  ${\*h_k}$. 

It is convenient to introduce the free energy
\begin{equation}
\label{freeenergy}
    \$F[\*h, \*S] = \sum_kg_k(h_k)  - \sum_{k\ne p}U_{kp}\*S_k\cdot\*S_p + \sum_k 2\*h_k\cdot\*S_k 
\end{equation}
where
\begin{align}
    g_k = -\frac{1}{\beta}\log&\big(1 + e^{-\beta(\xi_k +|\*h_k|)} + e^{-\beta(\xi_k -|\*h_k|)} \notag \\
    &+ e^{-\beta(2\xi_k+U_{kk})}\big) \label{eq:g-fun}.
\end{align}
We note that the mean-field decoupling excludes the self-interaction between the same $k$-eigenstate. The $U_{kk}$ term is treated exactly in evaluating the free energy of the doubly occupied $k$-state in $g_k$.
The different treatment of $k\ne p$ and $k = p$ arises because they have very different statistical properties which we shall discuss in the next section.

Minimizing the free energy with respect to $\*h_k$ and $\*S_k$, we obtain
\begin{equation}
    \frac{\pd g_k}{\pd \*h_k} = -2\*S_k \text{ and }\*h_k = \sum_{p(\ne k)}U_{kp}\*S_p\label{HF_MF_eqs},
\end{equation}
corresponding to our mean-field theory \eqref{mf2}. We shall see in the following that in the absence of disorder the interaction become constant $U_{kp}\rightarrow U_0$ and the mean-field equations reduce to the Stoner theory of ferromagnetism where a uniform solution $\*S_k\equiv\*S_0$ is obtained when $U_0$ satisfies the Stoner Criteria $U_0N(0)\ge 1$ \cite{Stoner_ori_Fazakas}, $N(0)$ is the density of states at the Fermi Surface. In the presence of disorder, $U_{kp}$ is no longer a constant and the solution $\*S_k$'s are in general different for different $k$'s. For strong enough disorder, the average of $\*S_k$ might vanish, resulting in a spin glass state as in classical spin models with random interaction between spins \cite{Spin_glass_ori, Repilca_trick1}. 

To study these possibilities, we have to compute the disorder-averaged free energy $\expval{\$F}_d$ as well as other physical observable $\expval{\$O}_d$, where $\expval{...}_d$ denotes the average over disorder configurations. We shall employ the replica trick \cite{Repilca_trick1} to compute the disorder averages in this paper.

\subsection{\label{sec:replica_trick}The Replica Approach}
We discuss how the replica trick is implemented in this section. We note that with our approximate free energy Eq.~\eqref{freeenergy} the disorder effects appears only through the interaction matrix elements $U_{kp}$ and $U_{kk}$.

To perform the replica trick, we assume that $U_{kp}, U_{k'p'}$ are mutually uncorrelated for $k(p)\neq k'(p')$ and follows a Gaussian distribution:
\begin{equation}
\label{pgaussian}
    P(U_{kp})\sim \frac{1}{\sqrt{2\pi\sigma^2}}e^{-\frac{(U_{kp} - \ovl{U})}{2\sigma^2}}
\end{equation}
where $\overline{U}$ is the mean and $\sigma^2$ the variance. We caution that this is a very crude approximation as all the effects of disorder is absorbed in only two parameters in our approximation. In particular, the distribution is $(k,p)$-independent and spatial information is lost, corresponding to infinite range interaction in $k$-space.

The parameters $U$ and $\sigma^2$ have been estimated for {\em localized} single particle wavefunction in $d$-dimension space where $|\phi_k(\*r_i)|^2\sim \frac{1}{L^d}e^{-\frac{|\*r_k-\*r_i|}{L}}$, where $L$ is the localization length. The details of the calculation can be found in ref.~\cite{Fermi_glass}. It was found that the mean and variance of $U_{kp} (k\neq p)$ satisfy:
\begin{gather}
    \expval{U_{kp}}_d \sim   \frac{U}{L^d}\times P_{kp}=\frac{U}{V} \label{esti_Ukp1} \\
    \expval{U_{kp}^2}_d - \expval{U_{kp}}^2_d \sim \frac{U^2}{L^{2d}}\times P_{kp}= \frac{U^2}{L^d V}\equiv\frac{h^2}{V} \label{esti_Ukp2}
\end{gather}
where $V$ is the system volume, and $P_{kp} \sim \frac{L^d}{V}$ signifies the probability of finding two localized states $(k,p)$ within a distance $\sim L$ to each other. We define $h^2=\sigma^2V=U^2/L^d$ for later convenience. Please note that both the mean and variance of the distribution are of order $\$O(1/V)$, which is necessary for obtaining finite result in thermodynamic limit \cite{SG_Nishimori}. In particular, we note that for extended states where $L^d\rightarrow V$, the effect of disorder vanishes in our approach, suggesting that a more refined approach beyond our crude approximation for $P(U_{kp})$ is needed in this case.

For $k=p$, we obtain
\begin{equation}
    \expval{U_{kk}}_d\sim\frac{U}{L^d}\label{esti_Ukk}
\end{equation}
as $P_{kk}=1$. It's important to realize that $\expval{U_{kk}}_d\sim \$O(1)$ whereas $\expval{U_{k\neq p}}_d\sim \$O(1/V)$. This important distinction suggests that $U_{kp}$ and $U_{kk}$ should be treated differently.We have treated $U_{kk}$ exactly in writing down $g_k$. We now employ the replica trick to treat the fluctuations in $U_{kp} (k\neq p)$ in the following.

The disorder averaged free energy is given by
\begin{align}
    -\beta\expval{\$F}_d = \int \prod_{k\ne p}\Big(\dif U_{kp}P(U_{kp})\Big)\log\$Z[U_{kp}]= \expval{\log\$Z}_d \label{Gaus_free_energy}
\end{align}
where $\$Z[U_{kp}]$ is the partition function for a given configuration of $\{U_{kp}\}$. The replica trick computes $\expval{\log\$Z}_d$ by using the identity
\begin{equation}
    \expval{\log\$Z}_d = \lim_{n\rightarrow 0}\frac{\expval{\$Z^n} - 1}{n}.
\end{equation}
For integer $n$, the integral over $dU_{kp})$ can be evaluated easily with the free energy \ (\ref{freeenergy}) and Eq.\ (\ref{pgaussian}) for $P(U_{kp})$.  As the distribution $P(U_{kp})$ is $(k,p)$-independent, the interaction term $\sum_{kp}U_{kp}\*S_k\cdot\*S_p$ is effectively infinite-range and a replica-mean-field theory can be employed in calculating $\expval{\$Z^n}_d$ \cite{Repilca_trick1}. We consider only the replica-symmetric solutions in this paper. The limit $n\rightarrow 0$ is taken afterward through an analytical continuation. The resulting mean-field equation is presented in the following with the mathematical details given in Appendix~\ref{App:replica_MF}. At zero temperature, we find that the replica symmetric (RS) solution is stable against replica symmetry breaking (RSB) terms. The detail of the stability analysis is presented in Appendix~\ref{App:RSB}.

After some lengthy algebra, we obtain the disorder-averaged free energy
\begin{widetext}
\begin{subequations}
\label{Free}
\begin{gather}
    \expval{f}_d=\frac{\expval{\$F}_d}{V} = 3\varphi_+\varphi_- +\frac{\*m^2}{4U} -\frac{1}{\beta V} \sum_k\int\frac{\dif \*y_k}{(2\pi)^{3/2}}e^{-\frac{\*y_k^2}{2}}\log\$Z_k[\*y_k] \label{replica_free_energy} \\
    \$Z_k[\*y_k] = \int\dif\*h_k\dif\*S_k e^{-\beta F_k[\*y_k, \*h_k, \*S_k]} \label{replica_free_energy_Z} \\
     F_k[\*y_k, \*h_k, \*S_k] = g_k(h_k)   - h\varphi_-\*S_{k}\cdot\*S_{k}-(\*m +\sqrt{2h\-\varphi}\*y_k-2\*h_k)\cdot\*S_{k} \label{replica_free_energy_F}.
\end{gather} 
\end{subequations}
\end{widetext}
where $\bar{\varphi}=\varphi_+-\frac{1}{2\beta}\varphi_-$. The free energy is characterized by three mean-field parameters $\*m$, $\varphi_+$ and $\varphi_-$ which are determined by the mean-field equations 
\begin{gather}
    M = \frac{m}{2U} =\frac{1}{V}\sum_k\expval{\expval{S_k^z}}_R \label{MF_eq01}\\
    \varphi_- = \frac{h}{3\sqrt{2h\-\varphi}}\frac{1}{V}\sum_k\expval{\expval{\*y_k\cdot\*S_k}}_R\label{MF_eq02}   \\
    \varphi_+ = \frac{h}{3}\frac{1}{V}\sum_k\expval{\expval{\*S_k\cdot\*S_k}}_R - \frac{h}{6\beta\sqrt{2h\-\varphi}}\frac{1}{V}\sum_k\expval{\expval{\*y_k\cdot\*S_k}}_R.  \label{MF_eq03}
\end{gather}
Here $M$ is the average spin polarization characterizes average magnetization ($g\mu_BM$) and we have chosen a particular direction for magnetization $\*M = M\hat z$ in writing down the equation for $M$. $\varphi_-\sim$ Edward-Anderson order parameter $q_{EA}$ \cite{Spin_glass_ori} which characterize spin-glass order while the first term in $\varphi_+$ characterizes the average spin amplitude. The bracket $\expval{\expval{\$O[\*S_k]}}_R$ is a shorthand of:
\begin{equation}
    \expval{\expval{\$O[\*S_k]}}_R = \int\frac{\dif\*y_k}{(2\pi)^{3/2}}e^{-\frac{\*y_k^2}{2}}\Big(\frac{1}{\$Z_k}\int\dif\*h_k\dif\*S_k \$O[\*S_k] e^{-\beta F_k}\Big)
\end{equation}
meaning that the configuration average over the $k$-state $\expval{\$O[\*S_k]}$ is performed first, followed by averaging over $\*y_k$. 

$\$Z_k$ and $F_k$ can be regarded as effective partition function and free energy for a given $k$-state. The replica-symmetric mean-field theory leads to the replacement of coupling between different $k$-states in Eq.~\eqref{freeenergy} by a random effective magnetic field $\*B_k = \*m +\sqrt{2h\-\varphi}\*y_k-2\*h_k$ that couples to individual spin $\*S_k$ and the disorder average is replaced by averaging over the Gaussian random variable $\*y_k = (y_k^x, y_k^y, y_k^z)$ that appears in $\*B_k$. 

At zero-temperature $\beta\rightarrow \infty$, the configuration average on state $k$ can be calculated by replacing $\expval{\$O[\*S_k]}$ with its saddle-point value $\$O[ \*S_k^{(0)}]$ in which $\*S_k^{(0)}$ minimizes the effective free energy $F_k$. This procedure recovers our mean-field result discussed above. A Landau theory is also developed from the free energy Eq.~\eqref{Free} to study the finite temperature phase transition. The mathematical details are shown in Appendices ~\ref{App:replica_MF} and ~\ref{App:GL}. The three self-consistent equations at zero-temperature are solved numerically and the result is presented in sec.~\ref{Sec:0T}. The corresponding finite temperature behaviors are presented in Sec.~\ref{Sec:GL}.

\section{Mean-Field results at zero temperature\label{Sec:0T}}
The numerical solutions of mean-field parameters at zero-temperature are presented in this section. We have considered the mean-field phase diagram in two dimensions $(d=2)$ for various values of $L, U$. For weak interaction, only states at the vicinity of Fermi surface are important. We assume a continues, structureless Fermi surface where the only important information enters our analysis is the electronic density of states and have expanded the density of states around the Fermi surface as
\begin{equation}
\label{DoS_expansion}
    N(\xi)\sim N(0)+\xi N'(0)+\frac{1}{2}\xi^2N''(0).   
\end{equation}
According to Eq.~\eqref{esti_Ukp1}-\eqref{esti_Ukk}, we have  set $U_{kk}\sim a\frac{h^2}{U}$ in solving the mean-field equations in which $a\sim\$O(1)$ is a positive factor and we have set $a=1$ in the following numerical calculation, without loss of generality. We use a typical data of density of states at Fermi surface $N(0)= 1.56 \ \rm{eV}^{-1}$ per electron spin from a monolayer Stoner ferromagnets \cite{Itin_EXP2D} and use a normalized Gaussian function to approximate its density of states and obtained the approximate data $N'(0)=-2.84\ \rm{eV}^{-2}$, $N''(0)=-23.23\ \rm{eV}^{-3}$ per electron spin. We note that a negative $N''(0)$ is required for Stoner magnets \cite{Coleman_many_body} and using other reasonable choices of density of states data affects the numerical results but does not affect our qualitative conclusions.

\begin{figure}[b]
\includegraphics[width=9cm]{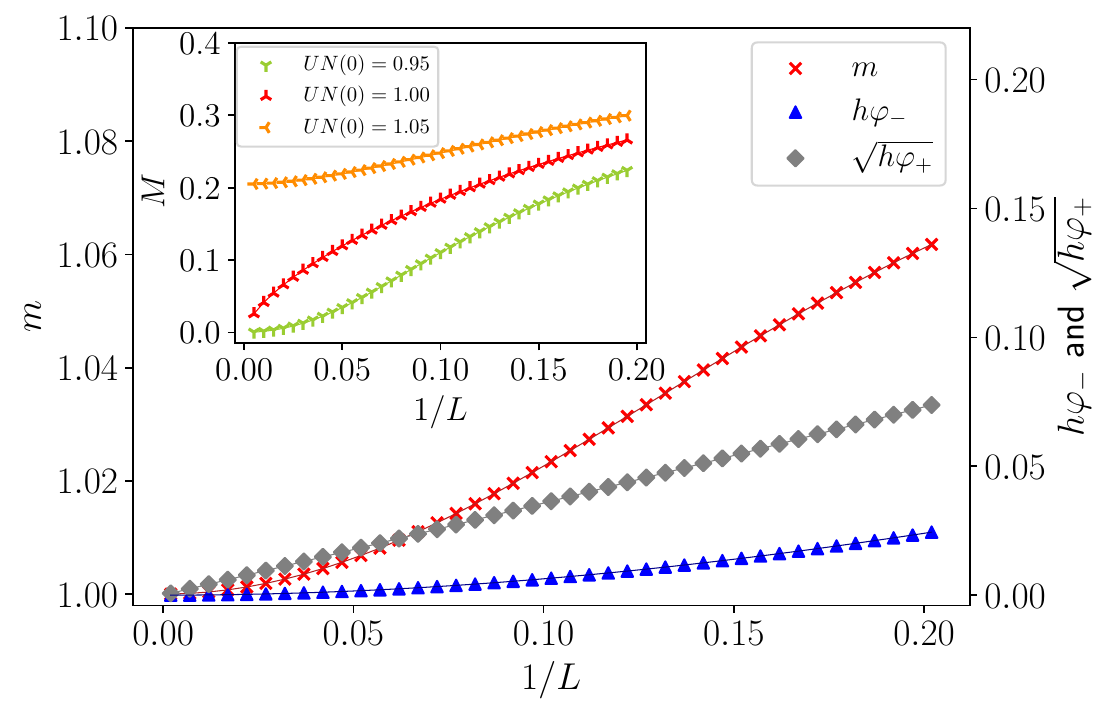}
\caption{This figure shows solutions for $UN(0)=1.2$, left axis shows values of $m=2UM$ and right axis shows values of $\sqrt{h\varphi_+}$ and $h\varphi_-$, scaled by $m_0$. The inset shows exact value of  magnetization $M$ for $UN(0)$ being close to 1.}
\label{fig:OT_sols01}
\end{figure}

We present in Fig.~\ref{fig:OT_sols01} the mean-field parameters $M,\varphi_-$ and $\varphi_+$ as a function of localization length $L$ (in unit of lattice spacing) for $UN(0) = 1.2$. Denote $M_0$ as the magnetization in clean limit, we show $m, h\varphi_-$ and $\sqrt{h\varphi_+}$ in unit of $m_0 = 2UM_0$ in this plot, keeping them in a same dimension.
We observe that the spin glass order parameter and spin amplitude parameter are both zero while magnetization are nonzero from weak localization regime and increase as the disorder strength increases. The coexistence of ferromagnetism and spin glass order (all three parameters are nonzero) is similar to classical spin glasses \cite{Repilca_trick1, Spin_glass_rev, SG_Nishimori} and represents a disordered ferromagnet. The enhancement in magnetization when the disorder strength increases is in agreement with previous experimental and theoretical results \cite{Itin_EXP4,Itin_THE1(perturbation)}.

In the limit $L^{-1}\rightarrow0$, the self-consistent equations Eq.~\eqref{MF_eq01}-\eqref{MF_eq03} can be studied analytically by expansion in powers of $h=U/L$ (2D). We found that in the leading order at $UN(0)>1$: 
\begin{equation}
    M = M_0 + \frac{h^2}{6U^2}\big(6a +5 - 2UN(0) \big)
\end{equation}
in which $M_0=\sqrt{\frac{6(UN(0)-1)}{-U^3N''(0)}}$ is the magnetization in the clean limit $h=0$. It shows that the magnetization is enhanced by disorder for moderate strength of interaction $UN(0)$. We caution, however that our result is not reliable in the regime $UN(0) \ll 1$, as the precise electronic band structure of the system will become important in determining $N(\xi)$, and our expansion around the Fermi surface  (Eq.~\eqref{DoS_expansion}) would not be valid.

The mean-field equations in the regime $UN(0)<1$ are also studied where several solutions of magnetization (exact values of $M$) around $UN(0)\sim 1$ are plotted in the inset of Fig.~\ref{fig:OT_sols01}. We find that magnetization can be induced by disorder even for $UN(0)<1$ as $L^{-1}$ increases.

\begin{figure}[b]
    \begin{subfigure}[ht]{0.49\linewidth}
         \includegraphics[width=4.4cm]{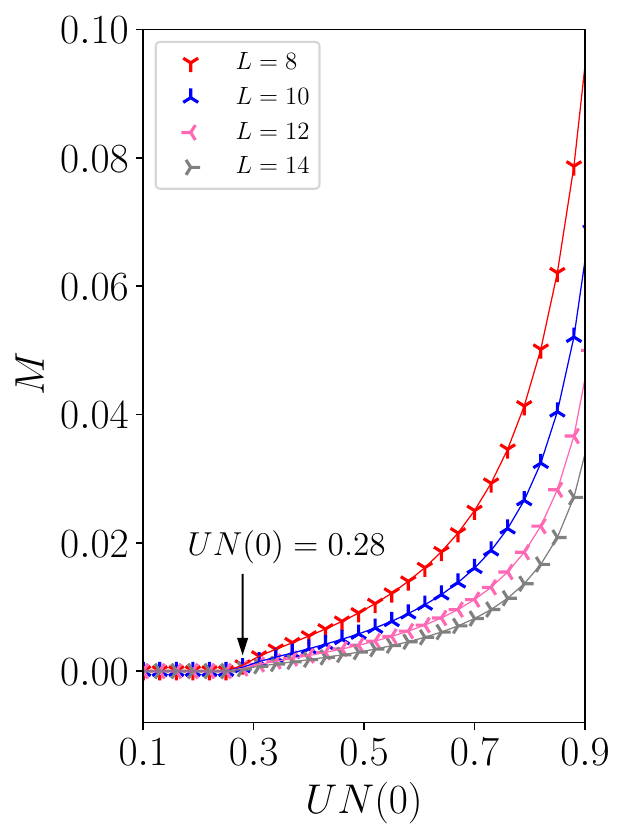}
        \caption{}
    \label{fig:OT_sols02-1}
    \end{subfigure}
    \begin{subfigure}[ht]{0.49\linewidth}
        \includegraphics[width=4.4cm]{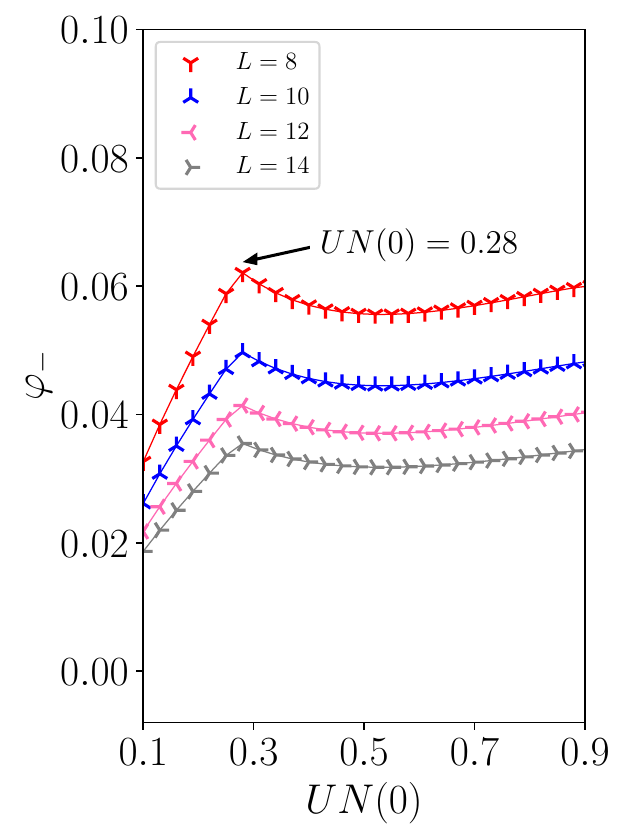}
        \caption{}
        \label{fig:OT_sols02-2}
    \end{subfigure}
\caption{The solutions for magnetization $M$ (a) and spin glass order $h\varphi_-$ (b) for a range of $UN(0)$ from 0.1 to 1.0 versus different localization length $L$. From top to bottom the localization length takes $8.0$ to $14.0$.}
\label{fig:OT_sols02}
\end{figure}

The regime $UN(0)\ll 1$ is interesting. Numerical solutions of the mean-field equations are shown in Fig.~\ref{fig:OT_sols02} in which the exact values of magnetization and spin-glass order parameter are plotted as a function of the $UN(0)$ for different values of localization length. 
We see that ferromagnetism develops at $UN(0)\ll1$ and increases with increasing disorder. From Fig.~\ref{fig:OT_sols02-1}, we see that magnetization is developed at $UN(0)\geq 0.28$ while from Fig.~\ref{fig:OT_sols02-2}, we see that the spin-glass order persists throughout the whole range of $UN(0)$ and a cusp appears in the solution for $\varphi_-$ at the phase transition point $UN(0)\sim0.28$.

The behavior of magnetism at small $UN(0)\ll1$ can be studied analytically in the weak localization limit $h^2\rightarrow0$. Using $U_{kk}\sim a\frac{U}{L^d}\sim a\frac{h^2}{U}$, we see that $U_{kk}\gg h^2N(0)$ in this limit suggesting we may solve the mean-field equation first in the limit $U_{kp}=0\ (k\neq p)$ while keeping $U_{kk}$ finite and then look at the effect of $U_{kp}$ perturbatively. At the starting limit, $U_{kp}=0$, it is easy to show that all states $k$ satisfying $\xi_k<U_{kk}/2$ are fully polarized but the directions of polarization are random when $U_{kp}=0$, i.e., a random magnetic moment state \cite{Fermi_glass} is obtained in this limit. To leading order in $U_{kp}$, the equation for magnetization is  (see Appendix~\ref{App:Expansion})
\begin{align}
    M &= \sqrt{\frac{16aUN(0)}{3\pi}}M\notag \\ &+\Big(\frac{N''(0)}{6}+\frac{N''(0)}{2\sqrt{\frac{h^2}{6}N(0)U_{kk}}}\sqrt{\frac{2}{\pi}}U_{kk}\Big)U^2M^3.
\end{align}
We have assumed $N''(0)<0$, which is the case for usual ferromagnetic materials. We note that $\sqrt{\frac{16aUN(0)}{3\pi}}\ll 1$ for small $UN(0)\ll1$ and magnetization $M=0$ in this limit. The system remains in a spin-glass state and magnetization develops only at $aUN(0)\geq0.59$, in qualitative agreement with what we observed in our numerical solutions.
\newline

\section{Finite temperature Landau Theory\label{Sec:GL}}
Phase transitions between paramagnetic and ferromagnetic and/or spin glass states are expected to take place at non-zero temperatures. In this section we derive the Landau theory from the disorder averaged free energy at temperature near the phase transitions and study various plausible phase transition between the three states as a function of disorder and interaction.
We see from Eq.~\eqref{replica_free_energy}-\eqref{replica_free_energy_Z} that $\expval{\$F}_d$ is a functional of $\*h_k$, $\*S_k$ as well as $M$, $h\varphi_-$ and $h\varphi_+$. To derive the Landau theory, We have to eliminate $\*h_k$, $\*S_k$ and $h\varphi_+$ using the mean-field equations, and expand $\expval{\$F}_d$ in powers of order parameters $M$ and $h\varphi_-$, noting that these parameters are small near phase transitions. We note that from Eq.~\eqref{MF_eq03} that $h\varphi_+$ is composed of $h\varphi_-$  and spin magnitude parameter $\sum_k\expval{\expval{\*S_k\cdot\*S_k}}_R$, which is not an order parameter representing a separate phase. The mathematical details of this straightforward but tedious calculation is presented in Appendix.~\ref{App:GL}. The resulting Landau free energy is of the form \cite{Spin_glass_expansion}
\begin{align}
   \$F_{\mathrm{L}} &= aM^2 + bM^4-cMh_{ext}  \notag \\
    & + \alpha(h\varphi_-)+ \gamma(h\varphi_-)^2 + \eta( M^2h\varphi_-) + ...\label{eq:Landau_FE}
\end{align}
in which $h_{ext}$ is the external magnetic field. Factors $a$, $b$, $c$ and $\alpha$, $\gamma$, $\eta$ are functions of temperature and  disorder strength $h$, expressed as:
\begin{align}
    &a = U^2\big(\frac{1}{U}-\frac{1}{V}\sum_k a_k\big),\ b = \frac{U^4}{V}\sum_k\frac{b_k}{2},\ c=\frac{U}{V}\sum_ka_k\notag \\
    &\alpha = \big[\frac{3}{2\beta V}\sum_ka_k-\frac{\frac{93h^2}{32\beta^2}\frac{1}{V}\sum_ka_k\frac{1}{V}\sum_kb_k}{1-\frac{h^2}{2V}\sum_ka_k^2+\frac{13h^2}{8\beta V}\sum_kb_k}\big]\notag \\
    &\gamma = \big[\frac{3}{4\beta V}\sum_ka_k^2 +\frac{93}{32\beta^2 V}\sum_kb_k\big],\ \eta = - \frac{U^2}{V}\sum_k\frac{9}{2\beta} b_k.
\end{align}
where the coefficients $a_k$ and $b_k$ are coming from the expansion (see Appendix.~\ref{App:GL} for detailed expressions)
\begin{equation}
    g_k\sim g_{0k} -a_k\*h_k\cdot\*h_k + \frac{1}{2}b_k(\*h_k\cdot\*h_k)^2 + ...\label{GL_gk}
\end{equation}

Minimizing the Landau free energy Eq.~\eqref{eq:Landau_FE} with respect to the order parameters $M$ and $h\varphi_-$, we obtain
\begin{subequations}
\begin{gather}
    \frac{\pd\$F_{\mathrm{L}}}{\pd M} = 0 \Rightarrow M^2 =-\frac{a+\eta (h\varphi_-)}{2b} \label{GL_MFQE01} \\
    \frac{\pd\$F_{\mathrm{L}}}{\pd (h\varphi_-)}=0\Rightarrow h\varphi_-=-\frac{\alpha + \eta M^2}{2\gamma}\label{GL_MFQE02}
\end{gather}
such that phase transitions, if exists, happen at $a+\eta(h\varphi_-)=0$ and/or $\alpha + \eta M^2=0$. The magnetic susceptibility above the ferromagnetic transition temperature is given by, 
\begin{equation}
\label{sus}
    \chi =\frac{1}{2} \frac{c}{a+\eta h\varphi_-}.
\end{equation}    
\end{subequations}
We observe that at the limit $h\rightarrow 0$ the magnetic susceptibility recovers the behaviors of pure Stoner magnets (see also Appendix.~\ref{App:GL}). The qualitative behavior of the system in the presence of $h\varphi_-\ne 0$ can be understood by assuming that the coefficients $b (>0)$, $c (>0)$, $\gamma (>0)$, and $\eta (<0)$ are temperature independent, with $a(T)=a_0(T-T_c)$ and $\alpha(T)=\alpha_0(T-T_f)$, where $T_c$ and $T_f$ are the ferromagnet and spin-glass transition temperatures, respectively, assuming that $T_f$ and $T_c$ are close to each other.

There are two possible scenarios with this approximation, $T_f>(<)T_c$. In the first case $T_f>T_c$, a spin-glass transition is observed at $T_f$, with both $h\varphi_-=0$ and $M=0$ at $T>T_f$, and $h\varphi_-\neq0$ and $M=0$ at $T_f>T>\overline{T}_c$, where $\overline{T}_c$ is determined by $a+\eta(h\varphi_-)=0$, or
\begin{subequations}
\begin{equation}
    \label{tran2}
    a_0(\overline{T}-T_c)-\eta h\frac{\alpha_0(\overline{T}-T_f)}{2\gamma}=0
\end{equation}
with solution
\begin{equation}
    \overline{T}_c=T_c-\frac{\eta h \alpha_0}{2\gamma a_0-\eta h \alpha_0}(T_f-T_c)>T_c
\end{equation}
\end{subequations}
as $\eta<0$, i.e., the ferromagnetic transition temperature is {\textit{ enhanced}} by he spin-glass order.  An analogous conclusion can also be reached for $T_c>T_f$. Notice that the magnetic susceptibility exhibits a cusp at temperature $T_f (>T_c)$ according to Eq.~\eqref{sus} because of the appearance of nonzero $h\varphi_-$ at $T<T_f$.  

To confirm the above results, we have computed the coefficients $a,b,c,\alpha,\gamma,\eta$ numerically to determine the phase boundaries at finite temperature for various values of $h^2$ and $UN(0)$. We observe that in agreement with the results at zero temperature, at finite temperature the ferromagnetism can also appear at $UN(0)\ll1$.

\begin{figure}
    \begin{subfigure}[ht]{0.49\linewidth}
        \centering
         \includegraphics[width=4.6cm]{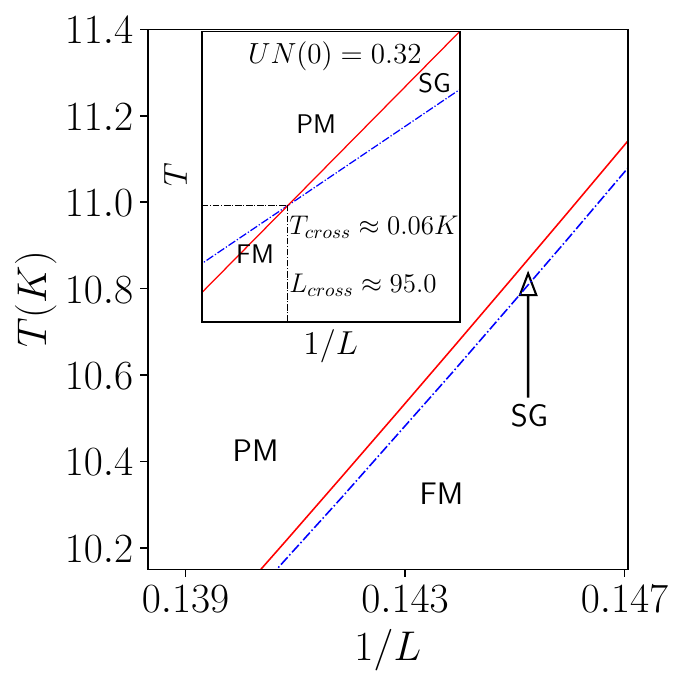}
        \caption{}
    \label{fig:GL01-1}
    \end{subfigure}
    \begin{subfigure}[ht]{0.49\linewidth}
        \centering
        \includegraphics[width=3.7cm]{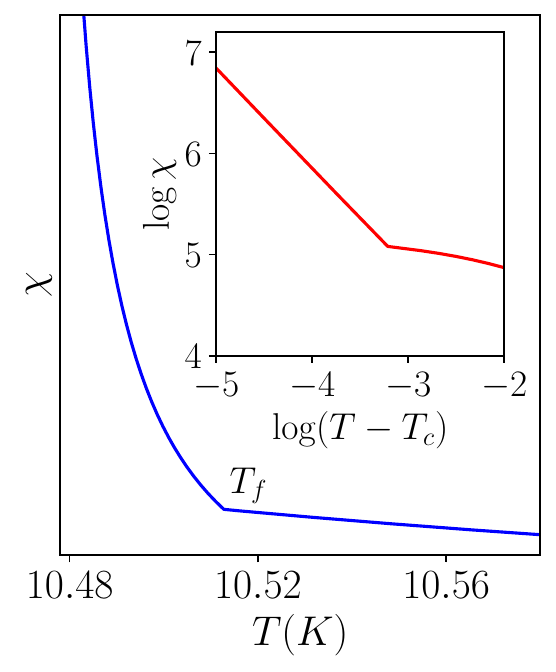}
        \caption{}
        \label{fig:GL01-2}
    \end{subfigure}
    \caption{Results for $UN(0)=0.32$. (a). The phase diagram, in which FM, SG and PM denote ferromagnetic phase, spin glass phase and paramagnetic phase; (b). Magnetic susceptibility $\chi(T)$ for $L=7.0$, in which $T_f$ denotes the temperature of spin glass transition.}
    \label{fig:GL01}
\end{figure}

The phase diagram at $UN(0)=0.32$ is shown in Fig.~\ref{fig:GL01-1}. The solid red line and dashed blue line represent the spin-glass and ferromagnet transitions, respectively. Both the spin-glass and ferromagnet transition temperatures are found to increase with disorder, with the ferromagnet transition temperature $T_c$ higher than the spin-glass transition temperature $T_f$ at weaker disorder (longer localization length) and with the order of transition reversed at stronger disorder (shorter localization length), in agreement with the general expectation that stronger disorder favors spin-glass state over ferromagnet. We note that the splitting between the spin-glass and ferromagnet transition temperatures is rather small ($\sim \$O(0.1K$)) at the regime $L\gg 1$ where our approximation for the density of states Eq.~\eqref{DoS_expansion} is valid. The magnetic susceptibly $\chi(T)$ at a strong disorder $L=7.0$ is shown in Fig.~\ref{fig:GL01-2}, the divergence at $T\rightarrow T_c$ and the cusp at $T\rightarrow T_f$ is clear. 

Further numerical calculations suggests that for small localization length (large $L^{-1}$ for strong disorder strength), a non-magnetic material ($UN(0)\ll 1$) can exhibit net ferromagnetism at low temperature, and for large localization length (small disorder strength) the ferromagnetic phase boundary is close to the clean limit results. In both cases of non-magnetic materials and ferromagnetic materials the Curie temperature can be strongly enhanced as the disorder strength increases.




\section{Discussion\label{Sec:Discussion}}

In this paper, we construct a replica-symmetric Stoner (RS-Stoner) mean-field theory to study how magnetization evolves in the presence of disorder in Stoner ferromagnet. To achieve this goal two crucial approximations are made: (i) We replace the usual on-site Hartree mean-field theory by a simplified mean-field theory where we do not have to determine the electronic eigenstates in the presence of disorder self-consistently. With this approximation the effects of disorder appear only in the effective interaction matrix element $U_{kp}$'s. (ii) To implement the replica trick, we make a very simple and crude approximation for the distribution of matrix elements $P(U_{kp})$ where we assume that $P(U_{kp})$ obeys a Gaussian distribution that depends only on two parameters. We also neglect the detail band structure of the material and assume that it is sufficient to expand the density of states around the Fermi surface Eq.~\eqref{DoS_expansion} in our mean-field theory analysis.
With these approximations a simple replica-symmetric Stoner mean-field theory can be constructed which can be analyzed rather easily. The aim of our paper is to see whether our calculation can qualitatively match the results from recent theoretical studies \cite{Itin_THE1(perturbation),Itin_THE5(Diagram),Itin_THE2(MF),Itin_THE2(MC),Itin_THE3(FP),Itin_THE4(FP),Itin_THE6(DMFT)} and to see whether new insights can be obtained.

We find that at zero-temperature the net magnetization $M$ is enhanced by disorder for moderate interaction strength $U$, including the situation $UN(0)<1$ which is non-magnetic in clean case, in agreement with recent theoretical work \cite{Itin_THE1(perturbation)}. For very weak interaction $UN(0)\ll1$ We predict a spin glass state can be found. However, we caution that interaction between far away spins is neglected in our simple approximation of $P(U_{kp})$, and a ferromagnetic state may still form at exponentially small temperature \cite{Fermi_glass}.

Our constructed Landau theory at finite temperature predicts that the Curie temperature is enhanced by disorder, consistent with results from other recent studies \cite{Itin_THE2(MF),Itin_THE6(DMFT)}. It also predicts that ferromagnetism can exist for $UN(0)\ll1$, in agreement with our zero temperature calculation and multiple experimental results \cite{Itin_EXP2,Itin_EXP3,Itin_EXP4}. in addition, it suggests that a narrow region of spin-glass state can exists between the paramagnetic high temperature state and ferromagnetic low temperature when disorder is strong enough. The existence of spin-glass state shows up as a discontinuity in the derivative of magnetic susceptibility against temperature at temperature above the ferromagnetic transition. 

Given the many approximations made in our theory, our results can at best be trusted only qualitatively. In particular, We expect that our Landau theory constructed from replica-symmetric solution is quantitatively not correct in the spin-glass state at finite temperature where replica-breaking terms become important, although it is expected that it can gives the phase boundaries correctly \cite{Repilca_trick1,Spin_glass_rev}. We note that the replica-symmetric solution considered in our theory is stable at zero-temperature, suggesting that the finite temperature replica-symmetric mean-field theory may still be qualitatively correct.

\begin{acknowledgments}
The project is supported by the School of Science, the Hong Kong University of Science and Technology.

\end{acknowledgments}
\newpage
\nocite{*}

\bibliography{apssamp}

\onecolumngrid
\appendix
\section{Details of the Replica mean-field Theory at Zero Temperature\label{App:replica_MF}}

We show here mathematical details of the replica mean-field theory at zero temperature. We begin with calculating $\expval{\$Z^n}_d$,
\begin{equation}
    \expval{\$Z^n}_d =\int\prod_{i = 1}^n\D\*S_{i}\D\bm h_{i}\prod_{k,p}\Big(\dif U_{kp}P(U_{kp})\Big)\Bigg[e^{-\beta\sum_i\Big(\sum_k g_k(h_k) + \sum_k2\*h_{ik}\cdot\*S_{ik}-\sum_{kp}U_{kp}(\*S_{ik}\cdot\*S_{ip})\Big)} \Bigg]
\end{equation}    
where $i$ is the replica index, $\D\*S_i\D\bm h_i = \prod_k\dif\*S_{ik}\dif\*h_{ik}$ and $P(U_{kp})=\sqrt{\frac{V}{2\pi h^2}}e^{-\frac{V(U_{kp} - U/V)}{2h^2}}$ where $V$ is the volume of the system. Integrate over $U_{kp}$ which are Gaussian integrals, we obtain:
\begin{equation}
    \expval{\$Z^n}_d =\int\prod_{i = 1}^n\D\*S_{ik}\D\*h_{ik}\Bigg[e^{-\beta\sum_i\Big(\sum_kg_k(h_k)+\sum_k2\*h_{ik}\cdot\*S_{ik}-\frac{U}{V}\sum_{kp}(\*S_{ik}\cdot\*S_{ip})\Big) +\frac{(h\beta)^2}{2V}\sum_{kp}(\omega_{kp})^2} \Bigg]
\end{equation} 
where $\omega_{kp} = \sum_i\Big(S^x_{ik}S^x_{ip} + S^y_{ik}S^y_{ip}+ S^z_{ik}S^z_{ip}\Big)$. We write
\begin{gather}
    \sum_{kp}(\omega_{kp})^2 = \sum_{\alpha\beta}\sum_{i\neq j}(\gamma_{ij}^{\alpha\beta})^2 +\sum_{i\alpha}\kappa^\alpha_i\kappa^\alpha_i + 2\sum_{\alpha\neq\beta}\sum_{i}\kappa^{\alpha\beta}_i\kappa^{\alpha\beta}_i \quad \text{for}\ \alpha,\beta = x, y, z \notag \\
    \gamma^{\alpha\beta}_{ij}=\sum_kS^\alpha_{ik} S^{\beta}_{jk},\quad \kappa^\alpha_i = \sum_k(S^\alpha_{ik})^2,\quad \kappa^{\alpha\beta}_i = \sum_kS^{\alpha}_{ik}S^{\beta}_{ik}\ \text{for}\ \alpha\neq\beta.
\end{gather}
and use the Hubbard-Stratonovich (HS) transformation to decouple $(\gamma^{\alpha\beta}_{ij})^2$, $\kappa^\alpha_i\kappa^\alpha_i$ and $\kappa^{\alpha\beta}_k\kappa^{\alpha\beta}_k$. Introducing the HS fields $q_{ij}^{\alpha\beta}$, $\chi^\alpha_i$ and $\chi^{\alpha\beta}_i$, we obtain:
\begin{align}
    & \ e^{\sum_{kp}\frac{(h\beta)^2}{2V}(\omega_{kp})^2} =\prod_{i\neq j}\int\prod_{\alpha\beta}\frac{\dif q^{\alpha\beta}_{ij} }{\sqrt{2\pi/V}}\int \prod_{\alpha}\frac{\dif\chi^\alpha_i}{\sqrt{2\pi/V}}\int \prod_{\alpha\neq\beta}\frac{\dif\chi^{\alpha\beta}_i}{\sqrt{2\pi/V}}\notag \\
    &\times \exp\Bigg[\sum_{i\neq j}\sum_{\alpha\beta}\big(h\beta q^{\alpha\beta}_{ij}\gamma^{\alpha\beta}_{ij}-\frac{V}{2}(q^{\alpha\beta}_{ij})^2\big)+ \sum_{i\alpha}\big(h\beta \chi^\alpha_i\kappa^\alpha_i-\frac{V}{2}(\chi^\alpha_i)^2\big) + \sum_{i,\alpha\neq\beta}\big(\frac{h\beta}{\sqrt{2}} \chi^{\alpha\beta}_i\kappa^{\alpha\beta}_i-\frac{V}{2}(\chi^{\alpha\beta}_i)^2\big)\Bigg]
\end{align}
We shall employ a saddle point approximation for the parameters $q_{ij}^{\alpha\beta}$, $\chi^\alpha_i$ and $\chi^{\alpha\beta}_i$ in the following as the system has infinite-range interactions in $k$-space. Assuming unbroken spin rotational symmetry, i.e., $q^{\alpha\beta}_{ij}\rightarrow q_{ij}\delta_{\alpha\beta}$, $\chi^\alpha_i\rightarrow\chi_i$ and $\chi^{\alpha\beta}_{ij}(\alpha\neq\beta)\rightarrow 0$. In this case, the n-replica partition function is 
\begin{align}
    \expval{\$Z^n}_d & \rightarrow\int\prod_{i\neq j}\D\*S_{ik}\D\*h_{ik} \Bigg[e^{-\beta\sum_i\Big(\sum_k g_k(h_{ik}) + \sum_k2\*h_{ik}\cdot\*S_{ik} - h\chi_i\*S_{ik}\cdot\*S_{ik}-\frac{U}{V}\sum_{kp}\*S_{ik}\cdot\*S_{ip}\Big)} \notag \\
    &e^{ +h\beta\big(\sum_{i\neq j}q_{ij}\sum_k\*S_{ik}\cdot\*S_{jk}\big)-\frac{3V}{2}\big(\sum_{i\neq j}q_{ij}^2 + \sum_i\chi^2_i\big)}\Bigg].
\end{align}
 Next we consider the term $\frac{\beta U}{V}\sum_{kp}\*S_{ik}\cdot\*S_{ip}\equiv\frac{\beta U}{V}(\sum_k\*S_{ik})^2$ where we perform another HS transformation:
\begin{align}
    \exp(\beta \frac{U}{V}\sum_i(\sum_{k}\*S_{ik})^2) = \int\prod_i\frac{\dif \*m_i}{(4\beta U\pi/V)^{3/2}}\exp(\sum_i\*m_i\cdot\sum_k\*S_{ik} - V\sum_i\frac{\*m_i^2}{4\beta U})
\end{align}
where $\dif \*m=\dif m_x\dif m_y\dif m_z$ and $\*m_i$ plays a role as an effective magnetic field. Treating $\*m_i$ by another saddle-point approximation, which corresponds to a Hartree mean-field theory. We obtain:
\begin{align}
    \expval{\$Z^n}_d \rightarrow \int\prod_{i\neq j}\D\*S_{ik}\D\*h_{ik} \Bigg[e^{-\beta\sum_{ik}\big(g_k(h_{ik}) + 2\*h_{ik}\cdot\*S_{ik} - h\chi_i\*S_{ik}\cdot\*S_{ik} + \*m_i\cdot\*S_{ik}\big)- \sum_i\frac{V\*m_i^2}{4\beta U}+h\beta\big(\sum_{i\neq j}q_{ij}\sum_k\*S_{ik}\cdot\*S_{jk}\big)-\frac{3V}{2}\big(\sum_{i\neq j}q_{ij}^2 + \sum_i\chi^2_i\big)}\Bigg].
\end{align}
Assuming replica-symmetric solutions: $q_{ij}= q$, $\chi_i = \chi_0$, $\*m_i = \*m_0$, the non-diagonal term $\sum_{i\neq j}q_{ij}\sum_k\*S_{ik}\cdot\*S_{jk}$ can eliminated by another HS transformation 
\begin{align}
   \quad e^{h\beta\sum_{i\neq j}q\sum_k\*S_{ik}\cdot\*S_{jk}}\rightarrow\quad e^{h\beta\sum_{i\neq j}q\sum_k\*S_{ik}\cdot\*S_{jk}} = e^{-h\beta\sum_iq\sum_k\*S_{ik}\cdot\*S_{ik}}\int\prod_k \frac{\dif \*y_k}{(2\pi)^{3/2}}e^{-\sum_k\frac{\*y_k^2}{2} + \sqrt{2\beta hq}\sum_{ik}\*y_k\cdot\*S_{ik}}.
\end{align}

Gathering all these HS fields, we obtain the n-replica symmetric partition function:
\begin{align}
    \expval{\$Z^n}_d &= \int\prod_k \frac{\dif \*y_k}{(2\pi)^{3/2}}\exp(-\sum_k\frac{\*y_k^2}{2} - \frac{3V}{2}n(n-1)q^2 -\frac{3V}{2}n\chi^2- \frac{nV\*m_0^2}{4\beta U}) \notag \\
    &\times \int\prod_{i}\D\*S_{ik}\D\*h_{ik} \Bigg[e^{-\beta\sum_{ik}\big(g_k(h_{ik}) + 2\*h_{ik}\cdot\*S_{ik} - h\chi_i\*S_{ik}\cdot\*S_{ik} + \*m_i\cdot\*S_{ik}\big)-h\beta\sum_iq\sum_k\*S_{ik}\cdot\*S_{ik}+ \sqrt{2\beta hq}\sum_{ik}\*y_k\cdot\*S_{ik}}\Bigg] \notag \\
    &= \int\prod_k \frac{\dif \*y_k}{(2\pi)^{3/2}}\exp(-\sum_k\frac{\*y_k^2}{2} - \frac{3V}{2}n(n-1)q^2 -\frac{3V}{2}n\chi^2- \frac{nV\*m_0^2}{4\beta U} + n\sum_k\log\$Z_{k}[\*y_k])\label{App_replica_Zn}
\end{align}
where we define 
\begin{align}
    \quad&\int\D\*S_{k}\D\*h_k e^{\sum_k\big(-\beta g(h_k) -2\beta \*h_k\cdot\*S_k + \*m_0\cdot\*S_{k} + \beta h(\chi_0-q)\*S_{k}\cdot\*S_{k}+ \sqrt{2\beta hq}\*y_k\cdot\*S_{k}\big)}  \notag \\
    & = \prod_k\int\dif\*S_{k}\dif\*h_k e^{-\beta g(h_k) -2\beta \*h_k\cdot\*S_k + \*m_0\cdot\*S_{k} + \beta h(\chi_0-q)\*S_{k}\cdot\*S_{k}+ \sqrt{2\beta hq}\*y_k\cdot\*S_{k}} = \prod_k \$Z_{k}[\*y_k].
\end{align}
From Eq.~\eqref{App_replica_Zn}, we can expand $\expval{\$Z^n}_d$ w.r.t. $n$ as:
\begin{equation}
    \expval{\$Z^n}_d = 1 + n\times \bigg[-\frac{3V}{2}(n-1)q^2 -\frac{3V}{2}\chi^2 -\frac{V\*m_0^2}{4\beta U} + \sum_k\int\frac{\dif \*y_k}{(2\pi)^{3/2}}e^{-\frac{\*y_k^2}{2}}\log\$Z_{0k}\bigg] + \$O(n^2)
\end{equation}
and the disorder averaged free energy density is obtained by applying the identity of replica trick:
\begin{gather}
    -\beta\expval{f}_d = \lim_{n\rightarrow 0}\frac{\expval{\$Z^n}_d-1}{Vn} = \frac{3}{2}(q^2 - \chi^2) -\frac{\*m_0^2}{4\beta U} + \frac{1}{V}\sum_k\int\frac{\dif \*y_k}{(2\pi)^{3/2}}e^{-\frac{\*y_k^2}{2}}\log\$Z_{k}\notag 
\end{gather}
Defining $\chi - q = \varphi_-$, $\chi+q = 2\beta\varphi_+$, $\-\varphi = \varphi_+-\frac{1}{2\beta}\varphi_- $ and $\*m = \frac{\*m_0}{\beta}$  we obtain 
\begin{gather}
    \expval{f}_d =  3\varphi_+\varphi_- +\frac{\*m^2}{4U} -\frac{1}{\beta V} \sum_k\int\frac{\dif \*y_k}{(2\pi)^{3/2}}e^{-\frac{\*y_k^2}{2}}\log\$Z_k[\*y_k] \notag\\
    \$Z_k = \int\dif\*h_k\dif\*S_k e^{-\beta\big( g(h_k) +2 \*h_k\cdot\*S_k -\*m\cdot\*S_{k} - h\varphi_-\*S_{k}\cdot\*S_{k}-\sqrt{2h\-\varphi}\*y_k\cdot\*S_{k}\big) }
\end{gather}
 The self-consistent equations for $\varphi_+$, $\varphi_-$ and $\*m$ are obtained by minimizing the free energy:
\begin{gather}
    \frac{\pd \expval{f}_d}{\pd \varphi_+} =0\quad\Rightarrow\quad3\varphi_- - \frac{h}{\sqrt{2h\-\varphi}}\frac{1}{V}\sum_k\expval{\expval{\*y_k\cdot\*S_k}}_R = 0   \notag \\
    \frac{\pd \expval{f}_d}{\pd \varphi_-} =0\quad\Rightarrow\quad 3\varphi_+ + \frac{h}{2\beta\sqrt{2h\-\varphi}}\frac{1}{V}\sum_k\expval{\expval{\*y_k\cdot\*S_k}}_R -  \frac{h}{V}\sum_k \expval{\expval{\*S_k\cdot\*S_k}}_R = 0\notag \\
    \frac{\pd \expval{f}_d}{\pd \*m} =0\quad\Rightarrow\quad\*m - 2U \frac{1}{V}\sum_k\expval{\expval{\*S_k}}_R = 0
\end{gather}
in which the average $\expval{\expval{...}}_R$ is defined as
\begin{gather}
    \expval{\expval{\$O[\*S_k]}}_R = \int\frac{\dif\*y_k}{(2\pi)^{3/2}}e^{-\frac{\*y_k^2}{2}}\Big(\frac{1}{\$Z_k}\int\dif\*h_k\dif\*S_k\$O[\*S_k] e^{-\beta F_k}\Big)\notag\\
    F_k = g_k(h_k) +2 \*h_k\cdot\*S_k  - h\varphi_-\*S_{k}\cdot\*S_{k}-(\*m +\sqrt{2h\-\varphi}\*y_k)\cdot\*S_{k}.
\end{gather}

We next derive the mean-field equations at the zero temperature. For simplicity we have assume that $U_{kk} = constant$ from now. We see that the interaction term $U_{kk}$ appears only in doubly occupied states in $g_k(h_k)$ and it is convenient to shift the single-particle energy $\xi_k\rightarrow \xi_k - U_{kk}/2$, obtaining at zero temperature:
\begin{equation}
    \lim_{\beta\rightarrow \infty}g_k(h_k) = \big(\xi_k-U_{kk}/2-|\*h_k|\big)\theta\big(-\xi_k+U_{kk}/2+|\*h_k|\big) + \big(\xi_k+U_{kk}/2+|\*h_k|\big)\theta\big(-\xi_k-U_{kk}/2-|\*h_k|\big).
\end{equation}
Moreover, $\-\varphi \rightarrow \varphi_+$ at zero temperature and
\begin{equation}
    F_{k} \rightarrow g_k(h_k) +2 \*h_k\cdot\*S_k  - h\varphi_-\*S_{k}\cdot\*S_{k}-(\*m+\sqrt{2h\varphi_+}\*y_k)\cdot\*S_{k}.
\end{equation}

At zero temperature we can calculate the average  $\expval{\expval{\$O[\*S_k]}}_R$ by replacing $\expval{\$O[\*S_k]}$ with the saddle-point value $\$O[\*S_k^{(s)}]$ where $\*S_k^{(s)}$ is given by minimizing the effective free energy $F_k$.
Minimizing it w.r.t. $\*h_k$ and $\*S_k$ we obtain
\begin{gather}
    2\*S^{(s)}_k = \big[\theta\big(-\xi_k + |\bm h^{(s)}_k| + U_{kk}/2\big) - \theta\big(-\xi_k - |\bm h^{(s)}_k|-U_{kk}/2\big)\big]\vec e_k \notag \\
    2\*h^{(s)}_k =[2h\varphi_-\*S_k^{(s)} + (\*m+\sqrt{2h\varphi_+}\*y_k)] = \big[2h\varphi_-S^{(s)}_k + |\*m + \sqrt{2h\varphi_+}\*y_k|\big] \vec e_k,\qquad 
\end{gather}
where $\vec e_k$ is the direction vector of the vector $\*Y_k = (\*m + \sqrt{2h\varphi_+}\*y_k)$. We note that $\*S^{(s)}_k\neq0$ in the region $|\xi_k|<U_{kk}/2$ even when $\bm h^{(s)}_k=0$, i.e., spin are polarized in this region even in the absence of magnetic field, which is a result first pointed out in ref.~\cite{Fermi_glass}. 

As long as $\*h_k^{(s)}\neq0$, the saddle-point solutions show that $\*S_k^{(s)}$ and $\*h_k^{(s)}$ are pointing at the same direction for all states $k$ ($\*Y_k\cdot\*S_k = Y_kS_k \cos\theta$ is minimized when $\cos\theta = 1$) and $S^{(s)}_k$ takes value $\frac{1}{2}$ for $-|\bm h^{(s)}_k|-U_{kk}/2<\xi_k<|\*h^{(s)}_k|+U_{kk}/2$. Choose $\hat z$ as the direction of magnetization $\*m=m\hat{z}$, we obtain for $m$:
\begin{equation}
    m_{\mathrm{tot}} = 2 U\sum_{\*y_k,k}\expval{S^z_k} = U\int\frac{\dif\*y_k}{(2\pi)^{3/2}}e^{-\frac{\*y_k^2}{2}} \int \dif\xi_k N(\xi_k)\big[\theta\big(-\xi_k + |\*h^{(s)}_k|+U_{kk}/2\big) - \theta\big(-\xi_k - |\*h^{(s)}_k|-U_{kk}/2\big) \big]\vec e_k\cdot\hat z \label{eq:M}.
\end{equation}
Using Eq.~\eqref{DoS_expansion}) for the density of states, we obtain
\begin{align}
    m &= U\int\frac{\dif \*y_k}{(2\pi)^{3/2}}e^{-\frac{\*y_k^2}{2}}\frac{m + \sqrt{2h\varphi_+}\*y_k\cdot\hat z}{|m\hat z + \sqrt{2h\varphi_+}\*y_k|}\Big\{\Big[N(0)U_{kk}+\frac{1}{24}N''(0)U_{kk}^3\Big]+\big(N(0)+\frac{1}{8}N''(0)U_{kk}^2\big)\Big[h\varphi_- \notag \\
    &+ |m\hat z + \sqrt{2h\varphi_+}\*y_k|\Big] +\frac{1}{8}N''(0)U_{kk}\Big[h\varphi_- + |m\hat z + \sqrt{2h\varphi_+}\*y_k|\Big]^2 + \frac{1}{24}N''(0)\Big[h\varphi_- + |m\hat z + \sqrt{2h\varphi_+}\*y_k|\Big]^3        \Big\}.\label{app:mfeq1}
\end{align}
In terms of $M = \frac{m}{2U}$ which is the physical magnetization,  this equation reduces to the equation from Stoner theory of ferromagnetism at $h=0$, $M = N(0) UM + \frac{N''(0)}{6}(UM)^3$. The Equations for $\varphi_+$ and $\varphi_-$ can be considered similarly, we obtain:
\begin{align}
    \varphi_+ &= \frac{h}{12}\int\frac{\dif \*y_k}{(2\pi)^{3/2}}e^{-\frac{\*y_k^2}{2}}\Big\{\Big[N(0)U_{kk}+\frac{1}{24}N''(0)U_{kk}^3\Big]+\big(N(0)+\frac{1}{8}N''(0)U_{kk}^2\big)\Big[h\varphi_- + |m\hat z + \sqrt{2h\varphi_+}\*y_k|\Big] \notag \\
    &+\frac{1}{8}N''(0)U_{kk}\Big[h\varphi_- + |m\hat z + \sqrt{2h\varphi_+}\*y_k|\Big]^2 + \frac{1}{24}N''(0)\Big[h\varphi_- + |m\hat z + \sqrt{2h\varphi_+}\*y_k|\Big]^3        \Big\} \label{app:mfeq2}
\end{align}
\begin{align}
    \varphi_- &= \frac{h}{6\sqrt{2h\varphi_+}}\int\frac{\dif \*y_k}{(2\pi)^{3/2}}e^{-\frac{\*y_k^2}{2}}\frac{m\hat z \cdot \*y_k+ \sqrt{2h\varphi_+}\*y_k^2}{|m\hat z + \sqrt{2h\varphi_+}\*y_k|}\Big\{\Big[N(0)U_{kk}+\frac{1}{24}N''(0)U_{kk}^3\Big]+\big(N(0)+\frac{1}{8}N''(0)U_{kk}^2\big)\Big[h\varphi_- \notag \\
    &+ |m\hat z + \sqrt{2h\varphi_+}\*y_k|\Big] +\frac{1}{8}N''(0)U_{kk}\Big[h\varphi_- + |m\hat z + \sqrt{2h\varphi_+}\*y_k|\Big]^2 + \frac{1}{24}N''(0)\Big[h\varphi_- + |m\hat z + \sqrt{2h\varphi_+}\*y_k|\Big]^3        \Big\}.\label{app:mfeq3}
\end{align}
The above equations are solved numerically to obtain the mean-field parameters $M$ and $\varphi_{\pm}$.

\section{Solution of mean-field equation at $UN(0)\ll 1$ at zero temperature}\label{App:Expansion}
In this section we analyze the mean-field equations \eqref{app:mfeq1}-\eqref{app:mfeq3} at the limit $UN(0)\ll 1$. Since  $U_{kk}\sim a\frac{h^2}{U}$, this limit is equivalent to $U_{kk}\gg h^2N(0)$. Thus we may first solve the mean-field equation at $U_{k\ne p}=0$ with $U_{k\ne p}$ being treated as perturbation. At zeroth order, $U_{k\ne p}=0$ and $\*m = 0$ by definition. We obtain the directional vector of saddle-point solution $\*S_k^{(s)}$ as
\begin{equation}
    \vec e_k = \frac{\*m+\sqrt{2h\varphi_+}\*y_k}{|\*m+\sqrt{2h\varphi_+}\*y_k|} \rightarrow \frac{\sqrt{2h\varphi_+}\*y_k}{\sqrt{2h\varphi_+}|\*y_k|} = \frac{\*y_k}{|\*y_k|}.
\end{equation}
In this situation, spins are fully polarized in the region $[-U_{kk}/2, U_{kk}/2]$ with fully random directions, i.e.,  a random magnetic moment state. Using the mean-field equation for $\varphi_+$, we obtain 
\begin{equation}
    h\varphi_+ = h^2\int_{-U_{kk}/2}^{U_{kk}/2}\dif\xi N(\xi)\frac{1}{12}\approx\frac{h^2}{12}N(0)U_{kk}
\end{equation}
which is proportional to $h^2N(0)$.  To first order in $U_{k\ne p}$, magnetization and spin glass order becomes nonzero. Denoting $\sqrt{2h\varphi_+}\sim\sqrt{\frac{h^2}{6}N(0)U_{kk}}\equiv C$, the directional vector of $\*S^{(s)}_k$ can be expressed as:
\begin{equation}
    \vec e_k=\frac{\*m + C\*y_k}{|\*m + C\*y_k|}\approx \frac{\*y_k}{|\*y_k|} + \frac{\*y_k^2\*m - (\*y_k\cdot\*m) \*y_k}{C|\*y_k|^3}\equiv\frac{\*y_k}{|\*y_k|}+\vec{\delta}_k(\*m) \label{small_m1}
\end{equation}
to first order of correction due to $\*m$ (higher order corrections are not necessary because mean field equation for $m$ is naturally cubic). Similarly, we have
\begin{equation}
    |\*m + C\*y_k|\sim C|\*y_k| + \frac{ \*y_k\cdot\*m}{|\*y_k|}\label{small_m2}.
\end{equation}
Using \eqref{app:mfeq1} and \eqref{app:mfeq3} for $m$ and $\varphi_-$, and replacing the directional vector $\vec e_k$ by \eqref{small_m1} and $|\*m +\sqrt{2h\varphi_+}|$ by \eqref{small_m2}, we obtain after performing the Gaussian integral over $\*y_k$ the spin glass order parameter to first order in $h^2N(0)$:
\begin{equation}
    h\varphi_-=\Big[\sqrt{\frac{4}{3\pi}h^2N(0)U_{kk}}+\frac{1}{2}h^2N(0)\Big]+\frac{h^2}{48}N''(0)U_{kk}^2 +\frac{\sqrt{3}}{36}\frac{hN''(0)U_{kk}^3}{\sqrt{\pi N(0)U_{kk}}},
\end{equation}
and the magnetization $m$,
\begin{equation}
    m = \sqrt{\frac{16aUN(0)}{3\pi}}m+\Big(\frac{N''(0)}{24}+\frac{N''(0)}{12C}\sqrt{\frac{2}{\pi}}U_{kk}\Big)m^3,
\end{equation}
where we have kept only lowest order term in $UN(0)$. Take $m = 2UM$, we have the equation for physical magnetization $M$ at $UN(0)\ll1$:
\begin{equation}
    M = \sqrt{\frac{16aUN(0)}{3\pi}}M+\Big(\frac{N''(0)}{6}+\frac{N''(0)}{2C}\sqrt{\frac{2}{\pi}}U_{kk}\Big)U^2M^3
\end{equation}
Notice that $\sqrt\frac{16aUN(0)}{3\pi}-1<0$ in the limit $UN(0)\ll1$ (for $a\sim \$O(1)$), $M=0$ in this limit.

\section{Details of Obtaining the Landau Free Energy\label{App:GL}}
We begin with the disorder averaged free energy, see Eq.~\eqref{replica_free_energy}, we first expand $g_k(h_k)$ in powers of $h_k$, assuming that $h_k$ is small near the phase transitions:
\begin{equation}
    g_k\sim g_{0k} -a_k\*h_k\cdot\*h_k + \frac{1}{2}b_k(\*h_k\cdot\*h_k)^2 + ...\label{GL_expansion}
\end{equation}
where $g_0 = -\frac{1}{\beta}\log(1 + 3e^{-\beta\xi_k} + e^{-\beta(2\xi_k + U_{kk})})$ and 
\begin{gather}
 a_k = \frac{\beta e^{-\beta\xi_k}}{1 + 2e^{-\beta\xi_k} + e^{-\beta(2\xi_k + U_{kk})}},\quad  b_k = -\frac{\beta^3}{6}\frac{e^{\beta(3\xi_k+ U_{kk})}(1+e^{\beta(\xi_k + U_{kk})}(e^{\beta\xi_k}-4))}{(e^{\beta\xi_k} + 2e^{\beta(2\xi_k+ U_{kk})} + e^{\beta(3\xi_k + U_{kk})})^2}.
\end{gather}
We then apply the mean-field equation $\frac{\pd g_k}{\pd \*h_k}=-2\*S_k$ to express $\*h_k$ in terms of $\*S_k$:
\begin{equation}
    \*h_k = \frac{\*S_k}{a_k - b_k\*h_k\cdot\*h_k}\approx\frac{\*S_k}{a_k} + \frac{\*S_k}{a_k^2}b_k(\*h_k\cdot\*h_k) \approx  \frac{\*S_k}{a_k}\big(1+ \frac{b_k}{a_k^3}\*S_k\cdot\*S_k\big).
\end{equation}
With the expansion, we can eliminate $\*h_k$ in the free energy, obtaining
\begin{gather}
    \expval{f}_d = 3\varphi_+\varphi_- + \frac{\*m^2}{4U} -\frac{1}{\beta V}\sum_k\int \frac{\dif\*y_k}{(2\pi)^{3/2}}e^{-\frac{\*y_k^2}{2}}\log\$Z_{k}\quad \\
    \$Z_{k}[\*S_k] =\int\dif\*S_k \ e^{-\beta F_k}= \int\dif\*S_k \ e^{-\beta\big[(\frac{1}{a_k} - h\varphi_-)\*S_k\cdot\*S_k + \frac{b_k}{2a_k^4}(\*S_k\cdot\*S_k)^2 - (\*m + \sqrt{2h\ovl\varphi}\*y_k)\cdot\*S_k\big]} .
\end{gather}
We then apply the saddle-point approximation:
\begin{equation}
    \frac{\delta \expval{f}_d}{\delta\*S_k} = 0 \Rightarrow \ovl{\*S}_k \approx  \frac{(\*m + \sqrt{2h\-\varphi} \*y_k)}{2(\frac{1}{a_k} - h\varphi_-)} - \frac{b_k}{2a_k^4}\frac{(\*m + \sqrt{2h\-\varphi} \*y_k)^3}{4(\frac{1}{a_k} - h\varphi_-)^4}\label{Saddle-point_S}
\end{equation}
where $\ovl{\*S}_k$ represents the saddle-point solution. The corresponding saddle-point free energy is
\begin{equation}
    \expval{f}_d \approx 3\varphi_+\varphi_- + \frac{\*m^2}{4U} +\frac{1}{V} \sum_k\int \frac{\dif\*y_k}{(2\pi)^{3/2}}e^{-\frac{\*y_k^2}{2}} F_k[\ovl{\*S}_k].
\end{equation}
To obtain the Landau free energy near phase transitions, we assume that $h\varphi_-$ is small near the phase transitions, and expand:
\begin{equation}
    \frac{1}{\frac{1}{a_k}-h\varphi_-}\approx a_k + a_k^2(h\varphi_-) + a_k^3(h\varphi_-)^2 + \$O\big((h\varphi_-)^3\big).
\end{equation}
On the other hand, using mean-field equation for $h\varphi_+$ Eq.~\eqref{MF_eq03} we also express it as powers of $h\varphi_-$ and $\*m$:
\begin{equation}
    h\varphi_+\sim\frac{\frac{h^2}{4\beta}\sum_ka_k+\frac{h^2}{3}\frac{\*m^2}{4}\Big(\sum_ka_k^2+2\sum_ka_k^3(h\varphi_-)-\frac{13}{4\beta}\sum_kb_k \Big)+\frac{13h^2}{16\beta^2}\sum_kb_k(h\varphi_-)-\frac{h^2}{4\beta}\sum_ka_k^3(h\varphi_-)^2}{1-\frac{h^2}{2}\sum_ka_k^2+\frac{13h^2}{8\beta}\sum_kb_k}+\$O(h^4).
\end{equation}
Substituting these expansions into the saddle-point solution $\ovl{\*S}_k$ Eq.~\eqref{Saddle-point_S} and performing the integral over $\*y_k$,  we obtain the Landau free energy:
\begin{align}
    \ovl{\$F}& \approx  M^2\Big[U-U^2\sum_ka_k-U^2\sum_k\frac{9b_k}{2\beta}h\varphi_-\Big] + M^4\Big[U^4\sum_k\frac{b_k}{2}\Big]\\
&+(h\varphi_-)\Big[\frac{3}{2\beta}\sum_ka_k-\frac{\frac{93h^2}{32\beta^2}\sum_ka_k\sum_kb_k}{1-\frac{h^2}{2}\sum_ka_k^2+\frac{13h^2}{8\beta}\sum_kb_k}\Big] \\
&+(h\varphi_-)^2\Big[\frac{3}{4\beta}\sum_ka_k^2 +\frac{93}{32\beta^2}\sum_kb_k\Big].
\end{align}
where we have substitute $M=\frac{m}{2U}$. Please note that this Landau free energy is only valid in the vicinity of phase transitions and contributions of factors are approximately linear in temperature ($\sim T -T_f$ and $\sim T-T_c$). Other contributions such as $\$O(h)^4$ in $h\varphi_-$ are neglected in this expansion. The factors of $\sum_k a_k^n$ can be evaluated numerically by energy integral
\begin{equation}
    \frac{1}{V}\sum_ka_k^n\approx\int\dif\xi a(\xi)^n\Big(N(0)+\xi N'(0)+\frac{1}{2}\xi^2N''(0) \Big).
\end{equation}
If an infinitesimal uniform magnetic field term $\sum_i \*h_{ext}\cdot \+c_{i\alpha}\bm\sigma_{\alpha\beta}c_{i\beta}$ is added to microscopic Hamiltonian Eq.~\eqref{original_model}, the magnetic field will appear in $\*h_k\rightarrow \*h_k + \*h_{ext}$ resulting in a linear term $ -\sum_k Ua_kM h_{ext}^z$ in the Landau free energy. We can see the longitudinal magnetic susceptibility can be expressed as:
\begin{equation}
    \chi = \frac{\sum_ka_k}{U\Big[\frac{1}{U}-\sum_ka_k-\sum_k\frac{9b_k}{2\beta}h\varphi_-\Big]}.
\end{equation}
In the limit of $h = 0$, the Landau free energy and magnetic susceptibility are reduced to the original form for Stoner ferromagnets.

\section{Stability of Replica-symmetric solution at Zero Temperature\label{App:RSB}}
We now show that at zero temperature the replica-symmetric solution is stable according to a replica-symmetry breaking (RSB) term. We shall write $q_{ij} = q+\delta q_{ij}$ in which $\delta q_{ij}$ is the small RSB term and see the related terms in the free energy. We begin with the n-replica partition function, Eq.~\eqref{App_replica_Zn}, write as:
\begin{equation}
    \expval{\$Z^n}_d = \Tr e^{-\beta H_R}
\end{equation}
where Trace denotes summing over all configurations of $\{\*S_{ik}\}$ and $\{\*h_{ik}\}$ as well as the disorder average over $\*y_k$. Write $q+\delta q_{ij}$, we can express $H_R$ as:
\begin{gather}
    H_R = \sum_{ik}\big(g_k(h_{ik}) + 2\*h_{ik}\cdot\*S_{ik} -h\varphi_-\*S_{ik}\cdot\*S_{ik} - (\*m + \sqrt{2h\varphi_+}\*y_k)\cdot\*S_{ik}\big) -h\sum_{k, i\ne j}\delta q_{ij}\*S_{ik}\cdot\*S_{jk}\notag \\
    + \sum_k\frac{\*y_k^2}{2\beta} + 3n\varphi_+\varphi_- + \frac{3n^2}{2\beta}q^2 + \frac{n\*m^2}{4U} + \frac{3}{2\beta}\sum_{i\ne j}\delta q_{ij}^2
\end{gather}
Because at zero temperature $\beta\rightarrow \infty$ the n-replica partition function is dominated by the configuration of $\{\*S_{ik}\}$ minimizing $H_R$, we need to minimize
\begin{equation}
    H_{\rm{eff}}[\*S_{ik}]=\sum_k\Big[\sum_{i}\big(g_k(h_{ik}) + 2\*h_{ik}\cdot\*S_{ik} -h\varphi_-\*S_{ik}\cdot\*S_{ik} - (\*m + \sqrt{2h\varphi_+}\*y_k)\cdot\*S_{ik}\big) -h\sum_{i\ne j}\delta q_{ij}\*S_{ik}\cdot\*S_{jk}\Big]
\end{equation}
and expand the solution of $\*S_{ik}$ as powers of $\delta q_{ij}$. To achieve this, we write $\*S_{ik} = \*S_k + \*x_{ik}$ in which $\*S_k$ is given by the replica-symmetric mean-field solution and $\*x_{ik}$ represents the contribution of the RSB term. Up to second order of $\*x_{ik}$, we obtain:
\begin{align}
    H_{\rm{eff}}[\*x_{ik}]&=\sum_k\Big[\sum_{i}\big(g_k(h_{ik}) + 2\*h_{ik}\cdot\*S_k -h\varphi_-\*S_k\cdot\*S_k - (\*m + \sqrt{2h\varphi_+}\*y_k)\cdot\*S_k\big) -\sum_ih\varphi_-\*x_{ik}\cdot\*x_{ik}\notag \\
    & -h\sum_{i\ne j}\delta q_{ij}\big(\*S_k\cdot\*S_k + \*S_k\cdot(\*x_{ik} + \*x_{jk}) + \*x_{ik}\cdot\*x_{jk}\big)\Big]
\end{align}
After minimizing $H_R$ w.r.t. $\*x_{ik}$, we have the solution:
\begin{equation}
    \*x_{ik} = \frac{\*S_k\sum_{j(\ne i)}\delta q_{ij}}{\varphi_- + \sum_{j(\ne i)}\delta q_{ij}} \approx 0 + \frac{\*S_k}{\varphi_-}\sum_{j(\ne i)}\delta q_{ij} - \frac{\*S_k}{\varphi_-^2} \sum_{j,l (\ne i)}\delta q_{ij}\delta q_{il} + \$O(\delta q^3)
\end{equation}
Substitute the solution into $H_{\rm{eff}}$, we find that:
\begin{align}
    &\text{to zeroth order of }\delta q_{ij}: \quad H_{\rm{eff}}^{(0)}=\sum_{ik}\Big[g_k(h_{ik}) + 2\*h_{ik}\cdot\*S_k -h\varphi_-\*S_k\cdot\*S_k - (\*m + \sqrt{2h\varphi_+}\*y_k)\cdot\*S_k\Big]\\
    &\text{to first order of }\delta q_{ij}: \quad H_{\rm{eff}}^{(1)}=-h\sum_{k,i\ne j}\delta q_{ij}\*S_k\cdot\*S_k = 0
    \\
    &\text{to second order of }\delta q_{ij}: \quad H_{\rm{eff}}^{(2)}=\sum_k-\frac{3h}{\varphi_-}\*S_k\cdot\*S_k\sum_{i, jl(\ne i)}\delta q_{ij}\delta q_{il}.
\end{align}
We can conclude that the at $\beta\rightarrow \infty$ the quadratic term of $H_R$ is 
\begin{equation}
    (-\frac{3h}{\varphi_-}\sum_k\*S_k^2)\sum_{i, jl(\ne i)}\delta q_{ij}\delta q_{il} = -z\sum_{i, jl(\ne i)}\delta q_{ij}\delta q_{il}
\end{equation}
in which $z\ge 0$ obviously. For the quadratic term $H_{\rm{eff}}^{(2)}$ being at least positive semi-definite, the replica-symmetric solution can be stable. We have checked the eigenvalues of the Hessian matrix at at zero temperature and they are given by $-z(n-2)$ and $-z(2n-2)$. At the limit $n\rightarrow0$ they are all non-negative. Our results are consistent with the well-known results of replica-symmetry breaking \cite{THE_REF_RSB}.

\end{document}